\documentclass[twocolumn,showpacs,preprintnumbers,prl,floatfix]{revtex4-1}
\usepackage{amssymb,amsmath,amsfonts}
\usepackage{graphicx,epsfig} % Include figure files
\usepackage{bm} % bold math

\begin{document}
\title{Functional modularity of background activities in normal and epileptic brain networks}
\author{M. Chavez,$^{1}$ M. Valencia,$^{1}$ V. Navarro,$^{2}$ V. Latora,$^{3,4}$
J. Martinerie,$^{1}$}
\affiliation{%
$^1$~CNRS UMR-7225, H\^{o}pital de la Salp\^{e}tri\`{e}re. 47~Bd. de l'H\^{o}pital, 75013 Paris, France}%
\affiliation{%
$^2$~Epilepsy Unit, H\^{o}pital de la Salp\^{e}tri\`{e}re, Paris, France}
\affiliation{%
$^3$~Dipartimento di Fisica e Astronomia, Universit\`a di Catania and INFN, Via S. Sofia, 64, 95123 Catania, Italy}
\affiliation{%
$^4$~Laboratorio sui Sistemi Complessi, Scuola Superiore di Catania, 
Via San Nullo 5/i, 95123 Catania, Italy}
%\date{\today}
%
\begin{abstract}
We analyze the connectivity structure of weighted brain networks extracted from spontaneous magnetoencephalographic (MEG) signals of healthy subjects and epileptic patients (suffering from absence seizures) recorded at rest. We find that, for the activities in the 5-14 Hz range, healthy brains exhibit a sparse connectivity, whereas the brain networks of patients display a rich connectivity with clear modular structure. Our results suggest that modularity plays a key role in the functional organization of brain areas during normal and pathological neural activities at rest. \end{abstract}
%\keywords{brain networks, MEG, complex networks, epilepsy}
\maketitle %

%%%%%%%%%%%%%%%%%%%%   INTRODUCTION  %%%%%%%%%%%%%%%%%%%%
From the brain to the Internet and to social groups, the
characterization of the connectivity patterns of complex systems has
revealed a wiring organization that can be captured neither by regular
lattices, nor by random graphs~\cite{boccaletti}. In
neurosciences, it is widely acknowledged that the emergence of several
pathological states is accompanied by changes in brain connectivity
patterns~\cite{varela01}. Recently, it has been found that functional
connectivity patterns obtained from magnetoencephalography (MEG) and
electroencephalography (EEG) signals during different pathological and
cognitive brain states (including epilepsy) display small-world (SW)
properties~\cite{functionalNetsSW}.
Empirical studies have also lead to the hypothesis that brain
functions rely on the coordination of a scattered mosaic of
functionally specialized brain regions (modules), forming a web-like
structure of neural assemblies~\cite{varela01}. Modularity
is a key concept in complex networks from RNA structures 
to social networks~\cite{modularityDefinition, Guimera2005}. 
A module is usually defined as a subset of units 
within a network, such that connections between them are denser than
connections with the rest of the network. In biological systems, it
is generally acknowledged that modularity results from evolutionary
constraints and plays a key role in robustness, flexibility and
stability~\cite{stabilityModules}.

Absence seizures are the most characteristic expression of non-convulsive generalized
epilepsy. Their main characteristic in brain signals is the occurrence of high amplitude, 
rhythmic spikeÐwave discharges synchronized over wide cortical areas, 
which manifests suddenly from a \textit{normal} background. Current
studies of brain connectivity mainly focus on the onset and evolution
of epileptic discharges. Nevertheless, little is known about the
organization of brain networks during spontaneous background
activities of patients, and the role of these connectivity patterns on
the emergence of absence seizures. In this Letter we study the {\em modular 
organization of brain networks} extracted from \emph{spontaneous} MEG 
signals of  epileptic patients and healthy subjects. The results of our 
analysis reveal a non-random structural organization in both normal
and pathological brain networks. In particular, the
functional networks of control subjects are characterized by a
sparse connectivity between the modules. In contrast, brain connectivity 
of epileptic patients (recorded out of seizures) displays a configuration 
where nodes in a functional module are connected to different functional modules. This modular configuration 
might play a key role in the integration of large scale brain activities, facilitating 
the emergence of epileptic discharges.

%%%%%%%%%%%%%%%%%%%%%%   DATA & METHODS  %%%%%%%%%%%%
The data used in this study were acquired from 5 healthy subjects and
5 epileptic patients suffering from absence seizures.  The study was
performed with written consent of the subjects and with the approval
of the local ethics committee. During the recordings, subjects and
patients were instructed to rest quietly, but alert, and keep their
eyes closed. The brain signals were acquired with a whole-head MEG
system (151 sensors; VSM MedTech, Coquitlam, BC, Canada), digitized at
$1.25$ kHz with a bandpass of $0-200$~Hz. All the analyses were performed on
338 non-overlapping quasi-stationary segments (206 for all the patients and 
132 for the healthy group) of 5 seconds without
eyes or muscular artefacts, nor epileptic activities (as, e.g., seizures or epileptic-like activity)
and far (at least 10 s) from recent epileptic discharge. 
In agreement with previous findings, surrogate data tests 
revealed that less than $4$~\% of interdependencies 
between the spontaneous brain activities were nonlinear~\cite{surrogate}. 
Thus, weighted brain networks were constructed by means of a definition of
functional links based on linear coherence. 
The squared modulus of the coherence between two time series $x_i(t)$ 
and $x_j(t)$ (normalized to zero mean and unit variance) was defined as:  
$|\Gamma_{ij}(f)|^2 = \frac{|S_{ij}(f)|^2}{S_{ii}(f)S_{jj}(f)}$, where
$S_{ii}$ and $S_{ij}$ (the spectral and cross-spectral densities) were 
estimated using the Welch's averaged periodogram method~\cite{brillingerBOOK}. 

Recent results show that correlations between 
magnetic fields sensors located at a distance less than $4$~cm 
can not distinguish between spontaneous activities of epileptic 
patients and control subjects~\cite{absenceSeizures}. 
To reduce the influence of these spurious correlations 
between MEG signals, we have excluded the nearest 
sensors (separated less than $5$~cm) from
the computation of coherence values. To perform the
statistical analysis of coherence values, we used Fisher's Z transform
of $\Gamma_{ij}$: $Z_{ij} = 0.5\ln \left(
  \frac{1+\Gamma_{ij}}{1-\Gamma_{ij}} \right)$. Under the hypothesis
of independence, $Z_{ij}$ has a normal distribution with expected
value 0 and variance $1/2N_b$, where $N_b$ is the number of
non-overlapping blocks used in the estimation of spectral densities~\cite{brillingerBOOK}. To correct for
multiple testing, the False Discovery Rate (FDR) method was applied to
each matrix of $\Gamma_{ij}$ values~\cite{FDR}. With this approach,
the threshold of significance $\Gamma_{th}$ was set such that the
expected fraction of false positives is restricted to $q \leq 0.01$.
Finally, in the network construction, a functional connection between
two of the $N=151$ nodes (brain sites) was assumed as an undirected
weighted link, i.e. we set the weight of the link between $i$ and $j$
as $w_{ij} = w_{ji} =\Gamma_{ij}$ when $\Gamma_{ij} > \Gamma_{th}$,
and $w_{ij}=0$ otherwise.

\begin{figure}[!htbp]
   \centering
%   \resizebox{0.8\textwidth}{!}{\includegraphics{figure1.eps}}
   \resizebox{0.95\columnwidth}{!}{\includegraphics{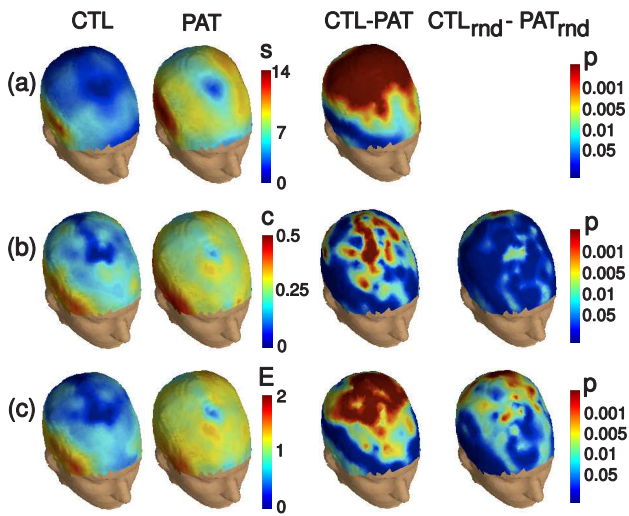}}
   \caption{Topographic distribution of node strength $s_i $ (a),
     weighted clustering coefficient $c_i$ (b), and efficiency $E_i$ (c).
     Control subjects (CTL), patients (PAT), and difference maps
     ($p$-values) of their comparison are reported for real and equivalent 
     random configurations.}  
   \label{topologyEpilepsy}
 \end{figure}

To characterize the network structure of healthy subjects and 
epilectic patients, we evaluated a list of measures for weighted graphs
\cite{boccaletti}. Namely, the node strength 
$s_i= \sum_k w_{ik}$, the weighted clustering
coefficient $c_i$~\cite{boccaletti}, the efficiency $E_i$ of a node $i$~\cite{latora}; 
as well as the respective averages over the graph nodes $S = 1/N
\sum_i s_i$, $C_w = 1/N \sum_i c_i$ and the global efficiency of the graph 
$E$.  As null model, brain
networks were compared to equivalent random graphs,  obtained by randomly 
rewiring the links of the original networks while keeping the same 
degree distribution \cite{Maslov2002}.  The statistical significance of a
given metric $\theta$ is assessed by quantifying its statistical
deviation from values obtained in the ensemble of randomized
networks. Let $\mu$ and $\sigma$ be the mean and SD of the parameter
$\theta$ computed from such an ensemble. The significance is given by
the ratio $\Sigma_{\theta}=|\theta - \mu|/\sigma$ whose p-value is
given by the Chebyshev's inequality~\cite{papulisBOOK}.  To assess significant differences
between topological features of the two groups, we used standard
nonparametric permutation methods, which account for spatial
dependences in the data~\cite{randomPermutat}. We used exhaustive
permutations ($10^5$) to estimate the empirical distribution under the
null hypothesis of no difference between the two groups. 

\begin{table}[!htb]
\begin{tabular}{cccccc}
\multicolumn1c{ }& \multicolumn1c{$S$}& \multicolumn1c{$C_w \ (\mu_C)$}&
\multicolumn1c{$E \ (\mu_E)$}& \multicolumn1c{$Q \ (\mu_Q)$} & \multicolumn1c{$N_{m}$} \cr \hline 
CTL & 4.04 & 0.233  (0.096) & 0.649  (0.892) & 0.538  (0.001) &  13.7  \cr 
PAT & 7.34 & 0.300  (0.160) & 0.893  (1.101) & 0.503  (0.006) &  8.30  \cr  \hline
\end{tabular}
\caption{Network properties in control subjects (CTL) and patients (PAT): 
mean node strength $S$, average weighted clustering coefficient $C_w$, 
global efficiency $E$, maximal modularity $Q$ and number of modules $N_m$. 
$\mu_\theta$ denotes the average of metric $\theta$ obtained from $20$ random graphs.}
\label{tableForNets}
\end{table}
Although we applied our approach to connectivity graphs obtained from
brain oscillations at the conventional frequency bands
($f<5$ Hz, $5<f<15$ Hz, $15<f<24$ Hz, $24<f<35$ and $f>35$ Hz), 
statistically significant differences between 
normal and epileptic brain networks were observed only for the  
brain activities in the extended alpha range ($5-14$~Hz).  Henceforth, all
results presented here refer to functional networks obtained at this frequency 
band. The basic network properties (averaged
over subjects) are summarized in Table~\ref{tableForNets}. The
structure of functional brain networks was found to be significatively
different from that of randomized counterparts. Namely, brain networks
of both patients and healthy subjects yielded a clustering
coefficient $C_w$ larger that that of randomized graphs ($p<10^{-3}$),
and values of efficiency as large as those of random graphs,  
indicating a small-world behavior. 
These results agree with previous findings suggesting
that brain sites have an optimal interaction with most
other brain regions~\cite{functionalNetsSW}. 
Furthermore, average network properties ($S$, $C_w$ and $E$) 
of epileptic patients were found to be
different ($p<10^{-3}$) from those of control subjects, supporting the
hypothesis that neural disturbances are correlated with changes in
functional network architectural features~\cite{varela01}.

A more detailed information on the differences between patients and
control subjects can be acquired by the analysis of the network at the
level of node properties. In Fig.~\ref{topologyEpilepsy} we report the
spatial distribution of node measures $s_i$, $c_i$ and $E_i$ (averaged
over all control subjects and over all patients) for each sensor of the
network.  Results indicate that epilectic patients have a richer node
connectivity than control subjects. Difference maps
clearly identify the centro-parietal regions as those brain areas with the
highest contrast between patients and control group. 
Although the node strength of epileptic patients is twice as large as 
that of healthy subjects, if links of real networks are randomly 
rewired by keeping the same degree distribution, no significant differences are observed 
between the two groups. This rules out the possibility that 
the difference in the number of connections  \emph{alone} 
could account for differences in $C_w$ and $E$. No significant
difference between healthy subjects and patients was observed by 
direct comparison of the spatial maps of the $\alpha$--activity power~\cite{supplementaryInfo}. 

A potential modularity of brain networks is suggested by the fact that
the networks display a clustering coefficient larger than that
obtained in random graphs~\cite{Ravasz2002}.
Previous studies over brain networks have 
used clustering methods to identify similar groups of 
brain activities. However, classical approaches such as those based on
principal components analysis (PCA) and independent components
analysis (ICA), make very strong statistical assumptions
(orthogonality and statistical independence of the retrieved
components, respectively) with no physiological justification~\cite{pcaICA}.
To find the network modules we have instead used an algorithm
based on a spectral embedding of graphs~\cite{lafon2006} (similar
theoretical frameworks have been recently proposed ~\cite{spectralMethods}). 
The algorithm is based on the definition of a Markov chain with a
transition probability matrix $P$ with entries $p_{ij}=
\frac{w_{ij}}{s_{i}}$. If $P^t$ is the
$t^\text{th}$ iterate of matrix $P$, the element $p_{ij}(t)$ encodes
the probability of moving from node $i$ to node $j$ through a random
walk of length $t$.  For an undirected and connected graph, the Perron Frobenius
theorem assures that $\mu_i^* = \frac{s_i}{\sum_k s_k}$ is the unique
stationary distribution of the Markov chain, such that $\lim_{t \to
 \infty} \sum_i p_{ij}(t)\mu_i (0) = \mu_j^* $.  For directed networks, 
 recent approaches are proposed to ensure this convergence~\cite{directedNets}. 

The random walk gives rise to a geometric diffusion with an associated
distance between nodes $i$ and $j$ defined as~\cite{lafon2006}:
$d^2_{ij}(t) =\sum_{k\geq 0}\frac{p_{ik}(t)-p_{jk}(t)}{\mu_k^*}$,
where the term $\mu_k^*$ is supposed to compensate for discrepancies
in local densities. By construction, distance between nodes is
strongly ruled by the connectivity of the graph, and it takes small
values if nodes are connected by many paths. Considering the spectral
representations of matrix $P$, one has a set of eigenvalues
$|\lambda_0| \geq |\lambda_1| \geq \ldots\geq |\lambda_{N-1}|$ and
eigenvectors $\varphi_k$ and $\psi_k$ such that $\varphi_k^TP =
\lambda_k \varphi_k^T $ and $P\psi_k=\lambda_k \psi_k$. The diffusion
distance can be written as: $d^2_{ij}(t) = \sum_{k \geq
  1}\lambda_k^{2t}(\psi_k(i) - \psi_k(j))^2$ where $\psi_k(j)$ denotes the 
component  $j$ of eigenvector $k$. The diffusion distance can be approximated 
(note that $\varphi_0= \mu^*$ and $\psi_0$ is a constant vector) to
a relative precision using the first $\beta$ nontrivial eigenvectors
and eigenvalues: 
$d^2_{ij}(t) \simeq \sum_{n = 1}^{\beta} \lambda_n^{2t}(\psi_n(i) - \psi_n(j))^2 $.  
This is equivalent to embed the graph in a low dimensional space
$\mathbb{R}^ \beta $, converting the diffusion distance between nodes
of the graph into Euclidean distance in $\mathbb{R}^ \beta $
~\cite{lafon2006}. This approach has the main advantage that it 
defines a meaningful representation of the graph and it leads 
to explicitly define a distance metric on the space
$\mathbb{R}^ \beta$ that reflects the connectivity of the network.
Graph modules are then extracted by a $k$-means
clustering algorithm in the embedding space. 
The algorithm starts with a random assignment of $N_m$ cluster
center. Then, the partition is updated by repeating the following
steps: \textit{i)} each point is assigned to the nearest center;
\textit{ii)} the new geometric center of each cluster are
recomputed. The algorithm stops when the assignments of nodes are no
longer changed. The algorithm reaches convergence in very few iterations after the
initial generations, but it only guarantees a convergence towards a
local optimum. In fact, the $k$-means is sensitive to the
initial choice of centroids. To overcome this drawback, we run the
algorithm several times (500) for different number of modules 
($N_m=2, \ldots , 15$) and return the partition yielding the largest
modularity $Q$. The modularity $Q(\mathcal{S})$ for a given partition $\mathcal{S}$ of 
a weighted network is defined as~\cite{modularityDefinition}:
$Q(\mathcal{S})=\sum_{s=1}^{N_m}{\left[{\frac{l_{s}}{L}-\left(
        \frac{d_{s}}{2L}\right)}^{2}\right]}$, 
where $L$ is the total weight of all connections in the
network, $l_{s}$ is the weight of links between vertices in
module $s$, and $d_{s}$ is the total weight of links in
module $s$. To select the optimal number of
diffusion coordinates, we have repeated the clustering algorithm for
all possible values of $\beta $, and looked for the largest value of
$Q$. 

\begin{figure}[!htb]
   \centering
%   \resizebox{0.75\textwidth}{!}{\includegraphics{figure2.eps}}
   \resizebox{0.9\columnwidth}{!}{\includegraphics{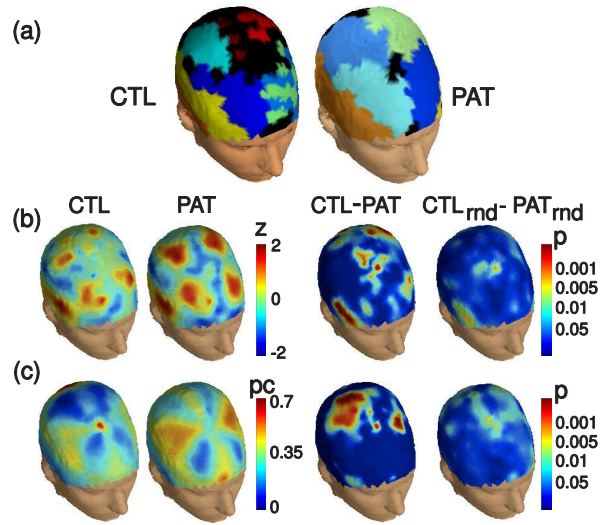}}
   \caption{(a) Brain sites belonging to each functional brain module
     were colored and superimposed onto an anatomical image (there is
     no color correspondence between the modules of patients and those
     of controls). For the sake of clarity, isolated nodes were
     colored in black. 
     Topographic distribution and difference maps ($p$-values) of 
     participation coefficient (b) and within-module degree (c) for real and equivalent random graphs.}
\label{modularityIndexMaps}
 \end{figure}
The method has been tested on synthetic networks \cite{performance},  
and used to find modules in brain networks. 
The presence of modules is actually
 confirmed by the high values of $Q$ obtained for the optimal
 partition of the brain networks of both control subjects and patients.
 The main result is that brain
 networks of epileptic patients have a structure consisting in a
 smaller number of modules ($N_m=8.30$) with respect to that of
 control subjects ($N_m=13.7$).  To assess the stability of the
 partition structure across healthy subjects and patients we used the
 adjusted Rand index $J$~\cite{randIndex}, which yields a normalized
 value between $0$ (if the two partitions are randomly drawn) and $1$
 (for an identical partition structure).  The values of $J$ 
indicate a high stability of the partition
 structure across all the epileptic patients ($J = 0.733$) and certain
 variability in the modular structure of control subjects ($J =
 0.479$).  

 Fig.~\ref{modularityIndexMaps}(a) illustrates the spatial
 distribution of the modules obtained for healthy subjects and
 patients.  Despite the differences, some modules fit well
 some known brain regions including occipital, parietal, and frontal
 areas.  Although a one-to-one assignment of anatomical brain areas to 
 the retrieved modules is difficult to define, modules assignment 
 provides the basis for the analysis of nodes according to their patterns of intra- and
 inter-modules connections~\cite{Guimera2005}. The within-module
 degree $z$-score measures how well connected the node $i$ is to other
 nodes in the same module, and is defined as $z_i =
 \frac{k_{is}-\overline{k}_{s}}{\sigma_{ks}}$; where $k_{is}$ is total
 weight of links of node $i$ to other nodes in its module $s$, while 
 $\overline{k}_{s}$ and $\sigma_{k_{s}}$ are average node strength and 
standard deviation of nodes in $s$. Thus node $i$ will display a large value 
of  $z_{i}$ if it has a large number of intra-modular connections relative 
to other nodes in the same module. The 
participation coefficient $pc_i = 1-\sum_{s=1}^{N_m}
 \left(\frac{k_{is}}{k_{i}}\right)^{2}$ quantifies instead to which
 extent a node $i$ is connected to different modules. This coefficient
 takes values of zero if a node has most of its links exclusively 
 with other nodes of its module, and $1$ if they are distributed among 
 different modules.
 
 In Figs.~\ref{modularityIndexMaps}(b-c) we report respectively the
 spatial distribution of the node participation coefficient $pc$ and
 of the $z$-score for control subjects and patients.  The difference
 map for the $z$-score reveals very small differences in the way the
 nodes are connected to other nodes in the same module. Conversely,
 the distribution of the participation coefficient $pc$ in control subjects strongly
differ from  that obtained for epileptic patients where nodes
 (specially those of the right centro-parietal areas) participate with their
 links in several modules. The same modular partition did not reveal
 significant differences between the equivalent random configurations 
 of both groups~\cite{supplementaryInfo}. This is a remarkable result as it supports
 the hypothesis that normal and pathological brain dynamics
 are characterized by different functional connectivity patterns~\cite{typeIerror}.
  
%%%%%%%%%%%%%%%%%%%   CONCLUSIONS   %%%%%%%%%%%%%%
In conclusion, in this paper we have addressed a fundamental problem
in neuroscience: characterizing the connectivity structure
of functional networks associated to normal and pathological neural dynamics.
From the analysis of spontaneous brain activities \emph{at rest}, we found that the architecture of
functional networks extracted from epileptic patients differs from  
that of healthy subjects. Interestingly, we identified a non-random modular 
structure of brain networks. Modularity analysis revealed that 
nodes of epileptic brain networks abnormally link different 
functional modules in the network. The connectivity of brain activities at 
the extended $\alpha$-band  of epileptic patients might play a putative role 
in the emergence of absence seizures. This leads up to the need of  
more refined studies, as in Ref~\cite{Srinivas2007}, to address the role 
of this architecture in the absences seizures.  

A modular description of brain networks might provide, more
in general, meaningful insights into the functional organization of
brain activities recorded with others neuroimaging techniques 
(EEG, MEG or fMRI) during diverse cognitive or pathological 
states. Applied to other multivariate data (e.g., financial or epidemiological time series), 
our approach could provide new insights
into the network structure of spatially extended systems. 

%%%%%%%%%%%%%%%%%%%   ACKNOWLEDGEMENTS   %%%%%%%%%%%%
The authors thank F.~Amor and Y.~Attal for kindly providing the data. 
This work was supported by the EU-GABA Project, contract no. 043309 (NEST).


\begin{thebibliography}{99}
\bibitem{performance} 
   The performance of the method has been tested on computer generated graphs 
   with tunable modularity structure. Even when half of a node's links points 
   to outside modules, the accuracy of the method is high ($\sim 80 \%$) 
   and comparable to that of the algorithm in Ref.~\cite{modularityDefinition}
\end{thebibliography}
\end{document}